\documentclass[preprint,pra,aps,showpacs]{revtex4}
\newcommand{\bb}{\mathbf}

\begin{document}
%

\title{Minimum dissipation principle \break in stationary 
non equilibrium states}

\author{L. Bertini}
\affiliation{Dipartimento di Matematica, Universit\`a di Roma La Sapienza
P.le A.\ Moro 2, 00185 Roma, Italy
}
\author{A. De Sole}
\affiliation{Departments of Mathematics, Harvard University, 
1 Oxford St., Cambridge, MA 02138, USA 
}
\author{D. Gabrielli}
\affiliation{
\mbox{
Dipartimento di Matematica, Universit\`a dell'Aquila} 
\break
67100 Coppito, L'Aquila, Italy} 
\author{G. Jona--Lasinio}
\affiliation{
Dipartimento di Fisica and INFN, Universit\`a di Roma La 
Sapienza P.le A.\ Moro 2, 00185 Roma, Italy
}
\author{C. Landim}
\affiliation{
IMPA, Estrada Dona Castorina 110, 
\mbox{J. Botanico, 22460 Rio de Janeiro, Brazil} 
\break
and 
CNRS UMR 6085, Universit\'e de Rouen,
76128 Mont--Saint--Aignan Cedex, France 
}



\begin{abstract}
We generalize to non equilibrium states Onsager's minimum dissipation
principle.  We also interpret this principle and some previous results
in terms of optimal control theory. Entropy production plays the role
of the cost necessary to drive the system to a prescribed macroscopic
configuration.
\bigskip
\bigskip\bigskip
\end{abstract}
%
%

%
%
\maketitle

\section{Introduction}
In his classic work on irreversible processes \cite{ONS1,ONS3}
Onsager introduced a quadratic functional of the thermodynamic fluxes
(time derivatives of thermodynamic variables) and an associated
variational principle, called the {\em minimum dissipation principle},
as a unifying macroscopic description of near equilibrium phenomena.
The evolution of the thermodynamic fluxes is given by a system of
linear equations which describes the relaxation to an equilibrium state.
This system is the Euler equation associated to the minimum
dissipation principle.  Moreover the value of this functional along
the solution of the relaxation equations is proportional to the
entropy production. The Onsager reciprocity relationships are
included in this principle which can be considered, as Onsager did, 
the dynamical analogue of the maximal entropy condition 
in equilibrium statistical mechanics.

Natural questions are how this theory can be extended to non
equilibrium stationary states and how the restriction to small
fluctuations can be relaxed.  Typical examples of non equilibrium
stationary states are systems in contact with reservoirs which
maintain a constant flow of some physical quantity (heat, matter,
electric charge,...).  Other examples are provided by systems in weak
external fields.
The rigorous study of simple stochastic models
of interacting particles
\cite{BDGJL1,BDGJL3,DLSlet,DLSas,DLSwas} has 
provided insight on these issues. 
These models typically give rise to non linear hydrodynamic equations 
and, for equilibrium states, Onsager's theory is recovered as a first
order approximation for small deviations.

In this paper we present a form of the minimum dissipation principle
valid for a wide class of models which includes stationary non
equilibrium states.
For a detailed discussion of our general assumptions, which are
satisfied for typical models of stochastic interacting particles, 
we refer to \cite{BDGJL1}. 
The basic points are a Markovian description of the microscopic
dynamics, the existence of a macroscopic dynamics (hydrodynamics) and
the validity of a dynamical large fluctuation principle which
generalizes the well known Boltzmann--Einstein formula for the
probability of thermodynamic fluctuations.  
The form of the hydrodynamic equations and the main consequence of the
large fluctuation principle, i.e.\ a Hamilton--Jacobi equation for the
entropy, will be introduced in the next section.
  
The minimum dissipation principle follows from the results obtained in
the previous papers \cite{BDGJL1} and in particular
from the decomposition of the hydrodynamic equation into two parts: 
a dissipative part proportional to the gradient of the entropy
and a non dissipative term orthogonal to it. 
The two terms have opposite transformation properties under time
reversal, the non dissipative part being in this respect akin to a
magnetic term. 

We emphasize that the minimum dissipation principle is of general
validity. This has to be contrasted with the principle of minimal
entropy production which is of limited validity \cite{ELS}.

\section{Basic macroscopic equations}\label{s:2}
We consider the macroscopic description of non equilibrium
conservative systems, that is systems in which the number of particles
is locally conserved.  The analysis of microscopic models of
stochastic lattice gases in contact with particles reservoirs and/or
weak external fields suggests that the following scheme may
characterize a wide class of physical systems.

\begin{enumerate}

\item
There exists an entropy functional $S(\{ \rho_i \})$ which depends on
a finite number of local thermodynamic variables $\rho_i$.  The
stationary state corresponds to a critical point of $S$ so that it is
a maximum of the entropy; here we stick to the physicists sign
convention for $S$ which is opposite to that of the previous papers.
The relationship between $S$ and the probability of a thermodynamic
fluctuation is given by the Boltzmann--Einstein formula
$P\approx\exp\{N^d \Delta S\}$ where $N^d$ is the volume of the
system.

\item 
The evolution of the fields $\rho_i=\rho_i(t,u)$, where $u$ represents
the macroscopic space coordinates and $t$ the macroscopic time, is
given by a divergence type hydrodynamical equations of the form (from now
on we omit for simplicity the indices)
\begin{equation}
\partial_t \rho  
=\frac 12 \nabla \cdot \big( D(\rho)\nabla \rho \big)
- \nabla \cdot \big( \chi(\rho) E  \big)
= \cal {D} (\rho)
\label{H}
\end{equation}
where $D$ is the diffusion matrix, $\chi$ the Onsager matrix, and $E$
the external field. 
Equation (\ref{H}) is obtained in the diffusive scaling with a
microscopic drift of order $1/N$.
The interaction with the reservoirs appears as boundary conditions to
be imposed on solutions of (\ref{H}).  
We expect  $D(\rho)$ and $\chi(\rho)$ to be local functions of the
thermodynamic variables. 
We assume that there exists a
unique stationary solution $\bar\rho$ of (\ref{H}), i.e.\ a profile
$\bar\rho(u)$, which satisfies the appropriate boundary conditions
such that ${\cal {D}} (\bar\rho)=0$. The entropy $S(\rho)$ attains its
maximum for $\rho=\bar\rho$ and we normalize it so that
$S(\bar\rho)=0$.

\item
The entropy $S$ satisfies the following Hamilton--Jacobi
functional derivative equation
\begin{equation}
\label{HJ}
\frac 12 \Big \langle \nabla \frac {\delta S}{\delta \rho} ,
\chi (\rho) \nabla \frac {\delta S}{\delta \rho} \Big \rangle
- \Big \langle \frac {\delta S}{\delta \rho} , 
{\cal {D} (\rho)}\Big\rangle = 0 
\end{equation}
where $\langle \cdot,\cdot\rangle$ means integration with respect to
the space variables.
\end{enumerate}

It is not obvious that a relationship such as (\ref{HJ}) connecting
the macroscopic entropy with hydrodynamics and Onsager matrix should
exist. In fact (\ref{HJ}) can be interpreted as a far reaching
generalization of the fluctuation--dissipation theorem since it allows
to express even in non equilibrium states a static quantity like the
entropy in terms of two dynamical macroscopic features of the system.
This is one of the main results of \cite{BDGJL1}. 

Another important result is the generalization to non equilibrium
states of the Onsager--Machlup theory of dynamical fluctuations
\cite{OMA}.
Suppose at time $t=-\infty$ the system is in the stationary state
$\bar\rho$ and at time $t=0$ we observe a profile $\rho$ of the
thermodynamic variable; we ask what is the trajectory along which this
fluctuation has been created.  
In equilibrium the Onsager--Machlup theory tells us that this
trajectory is the time reversal of the relaxation path which solves
the hydrodynamic equation (\ref{H}) with initial condition $\rho$.  In
non equilibrium the characterization of such a trajectory requires the
introduction of the adjoint hydrodynamics which is defined as the
hydrodynamic equation associated to the time reversed microscopic
dynamics.  Indeed in \cite{BDGJL1}, to which we refer for more
detail, we have shown that the exit path is the time reversed of the
solution to the adjoint hydrodynamics with initial condition $\rho$.

In the present paper we shall also present a non probabilistic point
of view in terms of optimal control theory.  
The dissipation function introduced in the next section is interpreted
as a cost function when an external perturbation is used to produce a
fluctuation.  The Hamilton--Jacobi equation becomes the Bellman
equation, see e.g.\ \cite{FR}, of the associated control problem.

\section{Minimum dissipation in non equilibrium states}
The extension of the minimum dissipation principle to non equilibrium
states follows from a structural property of the hydrodynamics in
non equilibrium states established in \cite{BDGJL1}.
By using the Hamilton--Jacobi equation (\ref{HJ}), we showed that we
can decompose the right hand side of (\ref{H}) as the sum of a
gradient of the entropy $S(\rho)$ and a vector field $\mathcal{A}$
orthogonal to it in the metric induced by the operator $K^{-1}$ where
$K f = - \nabla \cdot \big( \chi(\rho) \nabla f \big)$, namely
\begin{equation}
{\cal {D}} (\rho) = -{\frac {1}{2}} 
\nabla \cdot 
\Big( \chi(\rho) \nabla \frac {\delta S}{\delta \rho} \Big) 
+ {\mathcal{A}}(\rho)
\label{DEC}
\end{equation}
with
\begin{equation}
\Big\langle K\frac{\delta S}{\delta \rho} \, , K^{-1}\,
{\mathcal{A}}(\rho) \Big\rangle = \Big\langle \frac{\delta S}{\delta
\rho} \, ,\, {\mathcal{A}}(\rho) \Big\rangle = 0
\end{equation}
Since ${\cal A}$ is orthogonal to $\delta S/ \delta \rho$, it does
not contribute to the entropy production.

Both terms of the decomposition vanish in the stationary state that is
when $\rho = \bar \rho$.  Whereas in equilibrium the hydrodynamic is
the gradient flow of the entropy $S$, the term ${\mathcal{A}}(\rho)$
is characteristic of non equilibrium states.  Note that, for small
fluctuations $\rho\approx \bar\rho$, small differences in the chemical
potentials at the boundaries, small external fields $E$,
${\mathcal{A}}(\rho)$ becomes a second order quantity and Onsager
theory is a consistent approximation.

Equation (\ref{DEC}) is interesting because it separates out
the dissipative part of the hydrodynamic evolution associated
to the thermodynamic force $\frac {\delta S}{\delta \rho}$
and provides therefore an important physical information.
Notice that the thermodynamic force $\frac {\delta S}{\delta \rho}$
appears linearly in the hydrodynamic equation even when
this is non linear in the macroscopic variables.

In general, the two terms of the decomposition (\ref{DEC}) are non
local in space even if $\cal D$ is a local function of $\rho$. This is
the case for the simple exclusion process discussed in
\cite{BDGJL1,DLSlet}. 
Furthermore while the form of the hydrodynamic equation does not
depend explicitly on the chemical potentials, $\frac {\delta S}{\delta
\rho}$ and $\cal A$ do.

The adjoint hydrodynamic mentioned in Section \ref{s:2} is easily
obtained from the decomposition (\ref{DEC}). In fact in \cite{BDGJL1}
it is shown that it has the form 
\begin{equation}
\partial_t \rho =
{\cal {D}^*} (\rho) = -{\frac {1}{2}} 
\nabla \cdot 
\Big( \chi(\rho) \nabla \frac {\delta S}{\delta \rho} \Big) 
- {\mathcal{A}}(\rho)
\label{D*}
\end{equation}
that is $\cal A$ is odd under time reversal.

To understand how the decomposition (\ref{DEC}) arises microscopically
let us consider a stochastic lattice gas. Let $L$ be its Markov
generator;  we can write it as follows 
\begin{equation}
L=\frac {1}{2} (L + L^+) + \frac {1}{2} (L - L^+)
\label{HHe}
\end{equation}
where $L^+$ is the adjoint of $L$ with respect to the invariant
measure, namely the generator of the time reversed microscopic
dynamics.  The term $L -L^+$ behaves like a Liouville operator i.e.\
is antihermitian and, in the scaling limit, produces the term $\cal A$
in the hydrodynamic equation. 
This can be verified explicitly in the
boundary driven zero--range model discussed in \cite{BDGJL1}.  
In the appendix we briefly discuss this model under more general
conditions including a weak external field.
For this model
the decomposition (\ref{DEC}) can be obtained in two
ways: macroscopically from ${\cal D}(\rho)$ and $S(\rho)$,
microscopically deriving the hydrodynamics from the above
decomposition of the generator.

Since the adjoint generator can be written as $L^+= (L + L^+)/2 - (L -
L^+)/2$, the adjoint hydrodynamics must be of the form (\ref{D*}). 
In particular if the microscopic generator is
self--adjoint, we get ${\cal A}=0$ and thus ${\cal D}(\rho) ={\cal
  D}^*(\rho)$. On the other hand, it may happen that microscopic non
reversible processes, namely those for which $L\neq L^+$, can produce
macroscopic reversible hydrodynamics if $L -L^+$ does not contribute
to the hydrodynamic limit. This happens in the examples discussed in
\cite{GJL}.

To formulate the minimum dissipation principle, we want to construct a
functional of the variables $\rho$ and $\dot \rho$ such that the Euler
equation associated to the vanishing of the first variation under
arbitrary changes of $\dot \rho$ is the hydrodynamic equation
(\ref{H}).  We define the {\em dissipation function}
\begin{equation}
F(\rho , \dot \rho) = \Big\langle (\dot \rho - {\mathcal{A}}(\rho)) ,
K^{-1}(\dot \rho - {\mathcal{A}}(\rho))\Big\rangle
\label{DIS}
\end{equation} 
and the functional
\begin{equation}
\Phi (\rho , \dot \rho) = -\dot S(\rho) + F(\rho , \dot \rho) = -
\Big\langle {\frac {\delta S}{\delta \rho}}, \dot \rho \Big\rangle +
\Big\langle (\dot \rho - {\mathcal{A}}(\rho)) , K^{-1}(\dot \rho - {\mathcal{A}}(\rho))\Big\rangle
\label{VAR}
\end{equation} 
which generalize the corresponding definitions in \cite{ONS1,ONS3}.

It is easy to verify that
\begin{equation}
\delta_{\dot \rho} \Phi = 0
\label{EU}
\end{equation}
is equivalent to the hydrodynamic equation (\ref{H}).
Furthermore, a simple calculation gives
\begin{equation}
\label{FlI}
F |_{\dot \rho = \cal {D} (\rho)} = {\frac {1}{4}}
\Big \langle \nabla \frac {\delta S}{\delta \rho} \, ,\, 
\chi (\rho) \nabla \frac {\delta S}{\delta \rho} \Big \rangle
\end{equation}
that is $2F$ on the hydrodynamic trajectories equals the entropy
production rate as in Onsager's near equilibrium approximation. 

The dissipation function for the adjoint hydrodynamics is obtained by
changing the sign of ${\cal A}$ in (\ref{DIS}).

The decompositions (\ref{DEC}) and (\ref{D*}) remind the electrical
conduction in presence of a magnetic field. 
Consider the motion of electrons in a conductor: a simple 
model is given by the effective equation, see for example \cite{AM},
\begin{equation}
{\dot {\bf p}} = -e \Big( {\bf E} 
+ \frac{1}{mc} {\bf p} \wedge {\bf H} \Big)
- \frac{1}{\tau} {\bf p}
\label{ec}
\end{equation}
where ${\bf p}$ is the momentum, $e$ the electron charge, 
${\bf E}$ the electric field, ${\bf H}$ the magnetic field, 
$m$ the mass, $c$ the velocity of the light, and $\tau$ the relaxation
time. The dissipative term ${\bf p}/\tau$ is orthogonal to the
Lorenz force ${\bf p} \wedge {\bf H}$. 
We define time reversal as the transformation ${\bf p}\mapsto -{\bf p}$, 
${\bf H}\mapsto -{\bf H}$. The adjoint evolution is given by
\begin{equation}
{\dot {\bf p}} = e \Big( {\bf E} 
+\frac{1}{mc} {\bf p} \wedge {\bf H} \Big)
- \frac{1}{\tau} {\bf p}
\label{ec*}
\end{equation}
where the signs of the dissipation and the electromagnetic force
transform in analogy to (\ref{DEC}) and (\ref{D*}).

Let us consider in particular the Hall effect where we have conduction
along a rectangular plate immersed in a perpendicular magnetic field
$H$ with a potential difference across the long side.  The magnetic
field determines a potential difference across the short side of the
plate.  In our setting on the contrary it is the difference in
chemical potentials at the boundaries that introduces in the equations
a {\sl magnetic--like} term.

\section{Fluctuation, dissipation and optimal control}
The spontaneous motion of the systems  considered follows
the hydrodynamic equation (\ref{H}). We introduce an external
perturbation $v$ acting on the system in such a way that the 
hydrodynamical equation becomes
\begin{equation}
\partial_t \rho  
=\frac 12 \nabla \cdot \big( D(\rho)\nabla \rho \big)
- \nabla \cdot \big( \chi(\rho) E  \big)+v
={\cal {D}} (\rho) + v
\label{NH}
\end{equation} 
We want to choose $v$ to drive the system from its stationary
state $\bar\rho$ to an arbitrary state $\rho$ with minimal cost.
A simple cost function is
\begin{equation}
\frac{1}{2}\int_{t_1}^{t_2}\!ds\: \langle v(s), K^{-1}(\rho(s)) v(s) 
\rangle
\label{C}
\end{equation}
where $\rho(s)$ is the solution of (\ref{NH}) and we recall that
$K(\rho) f = - \nabla \cdot \big( \chi(\rho) \nabla f \big)$. 
More precisely, given $\rho(t_1)=\bar\rho$ we want to drive the system
to $\rho(t_2)=\rho$ by an external field $v$ which minimizes
(\ref{C}). 
This is a standard problem in control theory, see \cite{FR}.
Let 
\begin{equation}
{\cal V}(\rho)=\inf\: 
\frac{1}{2}\int_{t_1}^{t_2}\!ds\: \langle v(s), K^{-1}(\rho(s)) v(s) 
\rangle
\label{Vcost}
\end{equation}
where the infimum is taken with respect to all fields $v$ which drive
the system to $\rho$ in an arbitrary time interval $[t_1,t_2]$.  The
optimal field $v$ can be obtained by solving the Bellman equation
which reads
\begin{equation}
\min_{v}\Big\{ \frac{1}{2} 
\big\langle v, K^{-1}(\rho) v\big\rangle -
\Big\langle {\cal D} (\rho) + v, \frac{\delta {\cal V}}{\delta \rho} 
\Big\rangle
\Big\}=0
\label{Bel}
\end{equation}
It is easy to express the optimal $v$ in terms of ${\cal V}$; we get
\begin{equation}
v = K  \frac{\delta {\cal V}}{\delta \rho} 
\label{17}
\end{equation}
Hence (\ref{Bel}) now becomes
\begin{equation}
\frac{1}{2}
\Big\langle \frac{\delta {\cal V}}{\delta \rho},
K(\rho)  \frac{\delta {\cal V}}{\delta \rho} 
\Big\rangle 
+ 
\Big\langle {\cal D} (\rho), \frac{\delta{\cal V}}{\delta \rho} 
\Big\rangle
=0
\label{Bel2}
\end{equation}
By identifying the cost functional 
${\cal V}(\rho)$ with $-S(\rho)$, equation (\ref{Bel2})
coincides with the Hamilton--Jacobi equation (\ref{HJ}).

The time derivative of the entropy $S(\rho(t))$
when $\rho(t)$ is the solution of the hydrodynamic equation (\ref{H})
is proportional to the dissipation function (\ref{DIS}). 
By inserting the optimal $v$ (\ref{17}) in (\ref{NH}) and identifying
${\cal V}$ with $-S$, we get that the optimal trajectory $\rho(t)$
solves the time reversed adjoint hydrodynamics, namely
\begin{equation}
\partial_t \rho = - {\cal D}^*(\rho)
\label{h*-}
\end{equation}
where ${\cal D}^*(\rho)$ is given in (\ref{D*}).
The optimal field $v$ does not depend on the non dissipative part
$\cal A$ of the hydrodynamics.

The generalized Onsager--Machlup relationship \cite{BDGJL1}
states that a spontaneous fluctuation is created most likely by
following the solution of (\ref{h*-}).  By choosing the `right' cost
function (\ref{C}), this coincides with the best trajectory for
the control problem.

\appendix

\section{Zero range process with boundary conditions and weak external
field}

Consider the zero range process which models a nonlinear diffusion of
a lattice gas, see e.g.\ \cite{kl}.  The model is described by a
positive integer variable $\eta_{x}(\tau)$ representing the number of
particles at site $x$ and time $\tau$ of a finite subset $\Lambda_N$
of the $d$--dimensional lattice, $\Lambda_N = \mathbf{Z}^d \cap N
\Lambda$ where $\Lambda$ is a bounded open subset of $\bb{R}^d$.  The
particles jump with rates $g(\eta_x)$ to one of the nearest--neighbor
sites.  The function $g(k)$ is increasing and $g(0)=0$.  We assume
that our system interacts with particle reservoirs at the boundary of
$\Lambda_N$ whose activity at site $x$ is given by $\psi(x/N)$ for
some given smooth strictly positive function $\psi(u)$.  We consider
the process also in a (space dependent) external field $H(u)$ which
modifies the rates by a local drift.

The dynamics is specified by
the generator $L_{N}^H = 
L_{N,{\rm bulk}}^H+L_{N,{\rm bound.}}^H$ where
\begin{equation}\label{pgen}
\begin{array}{lll}
{\displaystyle 
L_{N,{\rm bulk}}^H f (\eta)} &=& {\displaystyle \frac 12 
\sum_{\scriptstyle x,y\in \Lambda_N \atop \scriptstyle |x-y|=1 } 
g(\eta_x) e^{H(y/N)-H(x/N)} 
\left[f(\eta^{x,y}) - f(\eta) \right]} 
\\
{\displaystyle 
L_{N,{\rm bound.}}^H f(\eta)} &=&{\displaystyle 
{\frac {1}{2}}
\sum_{x\in\Lambda_N, y \not\in\Lambda_N 
\atop |x-y|=1} \Big\{
g(\eta_x) e^{H(y/N)-H(x/N)}  
\left[ f(\eta^{x,-}) -f (\eta)\right]
}\\
&&\;\;\;\; {\displaystyle
+\, \psi(y/N) e^{H(x/N)-H(y/N)}  
[f(\eta^{x,+}) - f(\eta)]  \Big\} } 
\end{array}
\end{equation}
in which $\eta^{x,y}$ is the configuration obtained from $\eta$ when a
particle jumps from $x$ to $y$, and $\eta^{x,\pm}$ is the
configuration where we added (resp. subtracted) one particle at
$x$. With respect to the boundary driven zero range process considered
in \cite{BDGJL1}, we simply introduced in the macroscopic scale
a small space dependent drift $N^{-1} \nabla H(u)$ in the motion of
the particles.  The invariant measure $\mu_N$ of this model is a
product measure of the same form of the one discussed in \cite{BDGJL1}
in the case $H=0$.

Let
\begin{equation}
\label{Z=}
Z(\varphi) = 1 + \sum_{k=1}^{\infty}{\frac {\varphi^k}{g(1)
\cdots g(k)}}
\end{equation}
set
$R(\varphi) = \varphi  \, \frac {Z'(\varphi)}{Z(\varphi)}$,
and denote by $\phi(\rho)$ the inverse function of 
$\varphi \mapsto R(\varphi)$.
By the computations given in \cite{BDGJL1}, we get that the
hydrodynamic equation is 
\begin{equation}\label{H0RP}
\left\{
\begin{array}{l} 
{\displaystyle
\partial_t \rho (t,u) = \frac 12 \Delta \phi( \rho (t,u) ) - 
\nabla\cdot \left( \phi( \rho (t,u) ) \nabla H(u) \right)
\quad u \in \Lambda
}
\\
{\displaystyle \phi\left( \rho(t,u)  \right) = \psi(u) \quad u \in
\partial\Lambda} 
\end{array}
\right.
\end{equation}
which is of the form (\ref{H}) with $D(\rho)=\phi'(\rho)\openone$,
$\chi(\rho)=\phi(\rho)\openone$, and $E=\nabla H$.

Let us denote by $\bar\rho(u)$ the stationary solution of (\ref{H0RP}),
the associated activity $\lambda(u)=\phi(\bar\rho(u))$ solves 
\begin{equation}\label{H0RPst}
\left\{
\begin{array}{l} 
{\displaystyle
\frac 12 \Delta \lambda(u) - 
\nabla\cdot \left( \lambda(u)  \nabla H(u) \right) = 0
\quad u \in \Lambda
}
\\
{\displaystyle \lambda(u) = \psi(u) \quad u \in
\partial\Lambda} 
\end{array}
\right.
\end{equation}

It is not difficult to show that the entropy $S(\rho)$ is given by 
\begin{equation}
S(\rho)=
-\int_\Lambda \!du \: 
\left[ \rho(u) \log \frac {\phi(\rho(u))}{\lambda(u)}  
- \log \frac{Z(\phi(\rho(u)))}{Z(\lambda(u))}
\right]
\label{E}
\end{equation}
where $Z$ has been defined in (\ref{Z=}).

The decomposition (\ref{DEC}) of the hydrodynamics (\ref{H0RP}) is
\begin{equation}
\begin{array}{rcl}
{\displaystyle\vphantom{\Bigg\{}
-{\frac {1}{2}} \nabla \cdot 
\Big( \phi(\rho) \nabla \frac {\delta S}{\delta \rho} \Big) 
}
&=& 
{\displaystyle
\frac {1}{2} \Delta \phi(\rho) - \frac{1}{2} 
\nabla\cdot\Big( \frac{\phi(\rho)}{\lambda} \nabla \lambda
\Big) 
}\\
{\displaystyle
\vphantom{\Bigg\{}
\mathcal{A} (\rho) 
}
&=& 
{\displaystyle
-\nabla\cdot \left( \phi(\rho) \nabla H(u) \right) 
+ \frac{1}{2}
\nabla\cdot\Big( \frac{\phi(\rho)}{\lambda} \nabla \lambda\Big) 
}
\end{array}
\end{equation}
and it is not difficult to check that the orthogonality condition 
\begin{equation}
\Big\langle  \frac {\delta S}{\delta \rho} \,,\, \mathcal{A}(\rho)
\Big\rangle = 0
\end{equation}
holds.

Recalling that $\mu_N$ is the invariant measure let us consider the
process with initial condition $\nu$, where $\nu$ is a product measure
with density profile $\rho_0(u)$, namely 
$$
\frac{d\nu}{d\mu_N}(\eta) = 
\prod_{x\in\Lambda_N}  
\bigg(
\frac{\phi \big( \rho_0(x/N) \big)}{ \lambda(x/N) }  
\bigg)^{\eta(x)} 
\frac{Z\big( \lambda(x/N) \big)}{Z\big(\phi( \rho_0(x/N)) \big)}
$$
Let now $\nu_\tau$ be the distribution at time $\tau$ and 
${\cal H}(\nu_\tau|\mu_N)= 
- \int\!d\mu_N \frac{d\nu_\tau}{d \mu_N}\log \frac{d\nu_\tau}{d \mu_N}
$ the relative entropy of  $\nu_\tau$
with respect the $\mu_N$. By local equilibrium it is possible to show
that, as $N\to\infty$
\begin{equation}
\frac{d}{dt} {\cal H}(\nu_{N^2t}|\mu_N)
\to  \frac{1}{2} 
\Big \langle \nabla \frac {\delta S}{\delta \rho} (\rho(t))\, ,\, 
\phi (\rho(t)) \nabla \frac {\delta S}{\delta \rho} (\rho(t))\Big \rangle
\end{equation}
where $\rho(t)$ is the solution of the hydrodynamic equation 
(\ref{H0RP}) with initial condition $\rho_0$.
We see therefore that in the scaling limit we obtain the dissipation
in (\ref{FlI}) from the microscopic entropy production.

\acknowledgments 
We thank COFIN MIUR 2001017757 and 2002027798 for financial support.
This research was conducted by Alberto De Sole for the Clay Mathematics
Institute.

%
%

%
%


\begin{thebibliography}{99}



\bibitem{ONS1} L.\ Onsager, 
{\sl Reciprocal relations in irreversible processes. I.}
Phys.\ Rev.\ {\bf 37}, 405--426 (1931).
{\sl II.} Phys.\ Rev.\ {\bf 38}, 2265--2279 (1931).

\bibitem{ONS3}  L.\ Onsager, 
{\sl Theories and problems of liquid diffusion.}
Ann.\ N.Y.\ Acad.\ Sci.\ {\bf 46}, 241--265 (1945).

\bibitem{BDGJL1} 
L.\ Bertini, A.\ De Sole, D.\ Gabrielli, G.\ Jona--Lasinio, C.\ Landim,
{\sl Fluctuations in stationary non equilibrium states of irreversible
processes.} Phys.\ Rev.\ Lett.\ {\bf 87} (2001), 040601. 

{\sl Macroscopic fluctuation theory for stationary non equilibrium
state.}
J.\ Statist.\ Phys.\ {\bf 107}, 635--675 (2002).


\bibitem{DLSlet} B.\ Derrida, J.L.\ Lebowitz, E.R.\ Speer, 
{\sl Free energy functional for nonequilibrium systems:
an exactly solvable model.}
Phys.\ Rev.\ Lett.\ {\bf 87}, 150601 (2001).

{\sl Large deviation of the density profile in the steady state 
of the open symmetric simple exclusion process.}
J.\ Statist.\ Phys.\ {\bf 107}, 599--634 (2002).


\bibitem{BDGJL3} 
L.\ Bertini, A.\ De Sole, D.\ Gabrielli, G.\ Jona--Lasinio,
C.\ Landim,  
{\sl Large deviations for the boundary driven symmetric simple
exclusion process.} 
Math.\ Phys.\ Anal.\ Geom.\  (2003).


\bibitem{DLSas} B.\ Derrida, J.L.\ Lebowitz, E.R.\ Speer, 
{\sl Exact large deviation functional of a stationary 
open driven diffusive system: the asymmetric exclusion process.}
J.\ Statist.\ Phys.\ {\bf 110}, 775--809 (2003).



\bibitem{DLSwas} 
C.\ Enaud, B.\ Derrida, 
{\sl Large deviation functional of the weakly asymmetric exclusion
process.}  
Preprint arXiv:cond-mat/0307023 (2003).


\bibitem{ELS} G.L.\ Eyink, J.L.\ Lebowitz, H.\ Spohn, 
{\sl Microscopic
Origin of Hydrodynamic Behaviour: Entropy Production and the Steady
State}, in Proceedings of the Soviet/American Chaos Conference,
Woods Hole, MA, (1989).

\bibitem{OMA} 
L.\ Onsager, S.\ Machlup, {\sl Fluctuations and
irreversible processes.}\/  Physical Rev. {\bf 91} (1953), 1505--1512; 
Phys.\ Rev.\ {\bf 91} (1953), 1512--1515.

\bibitem{FR} W.H.\ Fleming, R.W.\ Rishel 
{\sl Deterministic and stochastic optimal control},
Springer 1975.

\bibitem{GJL}
D.\ Gabrielli, G.\ Jona--Lasinio, C.\ Landim,
{\sl Onsager symmetry from microscopic TP invariance.}
J.\ Statist.\ Phys.\ {\bf 96} (1999), 639--652.


\bibitem{AM}
N.W.\ Ashcroft, N.D.\ Mermin, {\sl Solid State Physics},
Holt--Saunders 1976.


\bibitem{kl} 
C.\ Kipnis, C.\ Landim, {\sl Scaling limits of interacting
particle systems.} Springer, Berlin 1999.


\end{thebibliography}
\end{document}